\documentclass[12pt]{iopart}
\begin{document}
\title{Octonionic Electrodynamics}
\author{M Gogberashvili}
\address{Andronikashvili Institute of Physics,
6 Tamarashvili St., Tbilisi 0177, Georgia }
\ead{gogbera@hotmail.com}
\begin{abstract}
Dirac's operator and Maxwell's equations in vacuum are derived in
the algebra of split octonions.  The approximations are given which
lead to classical Maxwell-Heaviside equations from full octonionic
equations. The non-existence of magnetic monopoles in classical
electrodynamics is connected with the using of associativity limit.
\end{abstract}
\maketitle


\section{Introduction}

Maxwell's equations, which harbor many beautiful mathematical
concepts, have been expressed in many forms since their discovery in
1873. Maxwell himself in his main book used the coordinate calculus
\cite{Maxwell}, however in the second edition included also
quaternionic representation. The original equations were a system of
16 equations, quaternionic and the familiar vector forms consist of
4 equations, and the application of bi-quaternions (or Clifford
algebras) results in a version of just one equation
\cite{q-Maxwell}. Still in the literature is absent octonionic form
of Maxwell's equation. It was already mentioned that the vector
algebra and Maxwell's equations should have connections with
octonions \cite{Si}. In this paper we want to show that in some
approximation classical Maxwell-Heaviside equations can be written
as the single continuity equation in the algebra of split octonions
over the reals.

Octonions form the widest normed algebra after the algebras of
real numbers, complex numbers, and quaternions \cite{Sc}. Since
their discovery, almost three decades before Maxwell's equations,
there have been various attempts to find appropriate uses for
octonions in physics (see reviews \cite{Oct}). One can point to
the possible impact of octonions on: Color symmetry \cite{Color};
GUTs \cite{GUT}; Representation of Clifford algebras \cite{Cliff};
Quantum mechanics \cite{QM}; Space-time symmetries \cite{Rel};
Field theory \cite{QFT}; Formulations of wave equations \cite{WE};
Quantum Hall effect \cite{Hall}; Strings and $M$-theory
\cite{String}, {\it etc}.

In our previous papers \cite{Go} the model where the geometry of
real world was described by the split octonions was introduced. In
\cite{Gog} the octonionic version of Dirac's equation was
formulated. In this paper, except of derivation of octonionic
Maxwell's equations in vacuum, we want to show that symbolic form of
Dirac's equations is just the result of the invariance of the
intervals in the octonionic geometry.


\section{Octonionic Geometry}

In the paper \cite{Go} real physical signals were associated with
the elements of split octonions,
\begin{equation} \label{s}
s = ct  + J_n x^n + j_n\hbar \lambda^n + Ic\hbar\omega  ~,~~~~~(n
= 1, 2, 3)
\end{equation}
where summing by the repeated indexes is performed. In (\ref{s})
the scalar unit is denoted as $1$, the three vector-like objects
as $J_n$, the three pseudo-vectors as $j_n$ and the pseudo-scalar
as $I$. The eight real parameters that multiply the basis units
denotes the time $t$, the special coordinates $x^n$, and some
quantities $\lambda^n$ and $\omega$ with the dimensions of
momentum$^{-1}$ and energy$^{-1}$ respectively. The line element
(\ref{s}) also contains two fundamental constants of physics - the
velocity of light $c$ and Planck's constant $\hbar$. The
appearance of these constants was connected with the existence of
two different classes of zero divisors in the algebra of split
octonions \cite{Go}.

The algebra of the basis elements of split octonions can be
written in the form:
\begin{eqnarray} \label{algebra}
J_n^2 = - j_n^2 = I^2 = 1~, \nonumber \\
J_nj_m = - j_mJ_n = - \epsilon_{nmk}J^k~,  \nonumber\\
J_nJ_m = - J_mJ_n = j_nj_m = -j_mj_n = \epsilon_{nmk} j^k~, \\
J_nI = - IJ_n = j_n~, \nonumber \\
j_nI = -Ij_n = J_n~, \nonumber
\end{eqnarray}
where $\epsilon_{nmk}$ is the fully antisymmetric tensor and
$n,m,k = 1,2,3$. From these formulas it is clear that to generate
a complete 8-dimensional basis of split octonions the
multiplication and distribution laws of only the three vector-like
elements $J_n$ are enough. The other two units $j_n$ and $I$ can
be expressed as binary and triple products
\begin{equation} \label{jI}
j_n = \frac{1}{2} \epsilon_{nmk}J^mJ^k~, ~~~~~I = J_nj_n
\end{equation}
(there is no summing in the second formula).

Using the conjugation rules of octonionic basis units
\begin{equation}
1^* = 1~, ~~~ J_n^* = - J_n~, ~~~ j_n^* = - j_n~, ~~~ I^* = - I~,
\end{equation}
one can find that the norm of (\ref{s}) (interval)
\begin{equation} \label{sN}
s^2 = s s^* = c^2t^2 - x_nx^n  + \hbar^2 \lambda_n\lambda^n -
c^2\hbar^2\omega^2 ~,
\end{equation}
has (4+4)-signature and in general is not positively defined.
However, as in the standard relativity we require
\begin{equation}
s^2 \geq 0~.
\end{equation}
In the classical limit $\hbar \rightarrow 0$ the expression
(\ref{sN}) reduces to the ordinary 4-dimensional formula for
space-time intervals.

Differentiating (\ref{s}) by the proper time $\tau$ the proper
velocity of a particle can be obtained
\begin{equation} \label{c}
\frac{ds}{d\tau} =  \frac{dt}{d\tau}\left[ c\left( 1 + I\hbar
\frac{d\omega}{dt}\right) + J_n\left(\frac{dx^n}{dt} + I\hbar
\frac{d\lambda^n}{dt}\right) \right] ~.
\end{equation}
Then the invariance of the norm (\ref{sN}) gives the relation
\begin{equation} \label{dtau/dt}
\beta = \frac{d\tau}{dt} = \sqrt{ \left[ 1 - \hbar^2
\left(\frac{d\omega}{dt} \right)^2\right] - \frac{v^2}{c^2} \left[
1 - \hbar^2\left(\frac{d\lambda^n}{dx^n}\right)^2\right] } ~,
\end{equation}
where
\begin{equation}
v_n = \frac{dx_n}{dt}
\end{equation}
is the 3-velocity. So the modified Lorentz factor (\ref{dtau/dt})
contains extra terms and the dispersion relation in the
(4+4)-space (\ref{sN}) has a form similar to that of
double-special relativity models \cite{double}.


\section{Symbolic Form of Dirac's Equation}

In \cite{Gog} the octonionic form of Dirac's equation, which in
some limit is equivalent to the standard one, was obtained. Here
we want to demonstrate simple derivation of Dirac's operator from
the condition of invariance of the octonionic interval (\ref{sN}).

For the observers with the time parameters $\tau$ and $t$ we can
write the relation
\begin{equation} \label{ds}
ds = \pm cd\tau = \pm cdt \beta~,
\end{equation}
where $\beta$ is expressed by (\ref{dtau/dt}). Dividing this
relation by $d\tau$ and multiplying it on the particle mass $m$ we
find:
\begin{equation} \label{dirac-s}
\frac{1}{\beta}\left[ mc\left( 1 + I\hbar
\frac{d\omega}{dt}\right) + J_n m \left(v^n + I\hbar
\frac{d\lambda^n}{dt}\right) \right] = \pm cm~.
\end{equation}
Let us assume that
\begin{equation} \label{lambda-A}
\frac{mc\hbar}{\beta}~ \frac{d\omega}{dt} = -\frac{e}{c}\varphi~,
~~~~~ \frac{m\hbar}{\beta}~ \frac{d\lambda^n}{dt} =
-\frac{e}{c}A^n~,
\end{equation}
where $\varphi$ and $A^n$ are components of the electro-magnetic
4-potential. By this assumption (\ref{sN}) takes a form similar to
intervals in Finsler-type theories with field-depended metrics.
For the reviews on Finsler theories see, for example
\cite{Finsler} and references therein.

Using the assumption (\ref{lambda-A}) the equation (\ref{dirac-s})
takes the form:
\begin{equation} \label{Dirac-s}
\left(\frac{\varepsilon}{c} - I\frac{e}{c}\varphi \right) +
J_n\left(p^n - I\frac{e}{c}A^n \right) \mp mc = 0~,
\end{equation}
where $\varepsilon = mc^2/\beta$ and $p^n = mv^n/\beta$ are energy
and momentum of the particle. This equation represents one of the
zero divisors in the algebra of split octonions. Importance of
zero divisors in physical applications of split algebras was
specially noted in \cite{So}.

The equation (\ref{Dirac-s}), which we receive from the invariance
of the interval (\ref{ds}), is the symbolic form of 4-dimensional
Dirac's equation. The role of four $\gamma$-matrices here is
played by the unit element of split octonions $1$ and the three
vector-like elements $J_n$. Instead of ordinary complex unit $i$
in (\ref{Dirac-s}) the basis element $I$ is used, and the factor
$\beta$ transforms to ordinary Lorentz formula if we use the limit
$\hbar \rightarrow 0$ in the definition (\ref{dtau/dt}).


\section{Maxwell's Equations in Vacuum}

The octonion that contains the electromagnetic potentials
$\varphi$ and $A_n$ we write as
\begin{equation} \label{A}
A = - \varphi + J_n A^n + j_n B^n + I b~, ~~~~~(n = 1,2,3)
\end{equation}
where $B^n$ and $b$ correspond to the extra degrees of freedom in
the octonionic algebra. Here we don't specify their meaning we only
want to obtain the approximations leading us to the classical
Maxwell-Heaviside equations that give successful explanation of most
experiments at low energies. Examples of problems in classical
electrodynamics where the fields $B^n$ and $b$ can play a role are:
magnetic monopoles \cite{Mon}, longitudinal electrodynamic force
\cite{Long}, the Abraham-Minkowski controversy \cite{Contro}, {\it
etc}.

To obtain week field approximation in octonionic equations let us
mention that, since we require positivity of norms, the elements of
split octonions should have hierarchial structure. This means that
the absolute value of the scalar element should be greater than
other elements and so on. From (\ref{jI}) it is also clear that the
pseudo-vector and pseudo-scalar units are secondary since they are
expressed by the fundamental vector-like elements. Appearance of
Planck's constant in the last two terms of (\ref{sN}) is another
indication that in classical limit we can neglect the values of
pseudo-vector and pseudo-scalar components. So it is natural to
consider that in (\ref{A})
\begin{equation}
|b|, |B^n| \ll |\varphi|, |A^n|~,
\end{equation}
and these components can be neglected. Invariance of octonionic
intervals then will guarantee that this inequality would be
preserved for different observers.

The octonionic differential operator we write as
\begin{equation} \label{nabla}
\nabla = \frac 1c \left(\frac{\partial}{\partial t} + I \frac
1\hbar \frac{\partial}{\partial \omega} \right)  +
J^n\left(\frac{\partial}{\partial x^n}+ I \frac 1\hbar
\frac{\partial}{\partial \lambda^n} \right) ~.
\end{equation}
Here we can also assume that influence of $\omega$ and $\lambda^n$
can be ignored in the classical limit. The norm of $\nabla $ when
fields don't depend on $\omega$ and $\lambda^n$ is the ordinary
4-d'Alembertian.

Assuming in (\ref{A}) that $B^n$ and $b$ are small (or are
constants) and $A^n$ and $\varphi$ are independent of $\omega$ and
$\lambda$, we can define the electro-magnetic field in the form:
\begin{equation} \label{F}
\nabla A = F = \left(-\frac{1}{c}\frac{\partial \varphi}{\partial
t} + \frac{\partial A^n}{\partial x^n} \right) + J_n  E^n + j_n
H^n ~, ~~~~~(n = 1,2,3)
\end{equation}
where
\begin{equation}
E^n = \frac 1c \frac{\partial A^n}{\partial t} - \frac{\partial
\varphi}{\partial x_n} ~, ~~~~ H^n = \epsilon^{nmk} \frac{\partial
A_k}{\partial x^m}~,
\end{equation}
are components of 3-vectors of electric and magnetic fields
respectively.

Then we can postulate the Lorenz gauge (derived by L. Lorenz in
1867,  and not by H. A. Lorentz, as refereed in some modern papers)
\begin{equation} \label{Lorenz}
\frac{1}{c}\frac{\partial \varphi}{\partial t} - \frac{\partial
A^n}{\partial x^n} = 0~,
\end{equation}
or weaker condition where zero in (\ref{Lorenz}) is replaced by a
constant, and write continuity equation as the product of the
octonions (\ref{nabla}) and (\ref{F}),
\begin{eqnarray} \label{nabla-F}
\nabla F =  \frac{\partial E^n}{\partial x^n} + J_k
\left(\frac{1}{c} \frac{\partial E^n}{\partial t} -
\epsilon^{nmk}\frac{\partial H_m}{\partial x^n}\right) + \nonumber
\\
+j_k \left( \frac 1c \frac{\partial H^n}{\partial t} +
\epsilon^{nmk}\frac{\partial E_m}{\partial x^n}\right) + I
\frac{\partial H^n}{\partial x^n} = 0~.
\end{eqnarray}
Different signs in the second and third terms of this equation are
the result of the using of the algebra (\ref{algebra}), in
particular
\begin{equation}
J_nj_m = - \epsilon_{nmk}J^k~, ~~~~~ J_nJ_m = \epsilon_{nmk} j^k~.
\end{equation}
Equating to zero coefficients in front of the four octonionic
basis units in (\ref{nabla-F}) results the full set of the
homogeneous Maxwell's equations.

We can write also octonionic current function in the form:
\begin{equation} \label{current}
\varrho = \rho + J_n\frac 1c \sigma^n~,
\end{equation}
where $\rho$ is the electric charge density and $\sigma^n$ are the
components of electric current vector. As before we ignored
pseudo-vector and pseudo-scalar parts in (\ref{current}).

Finally we can write the complete set of inhomogeneous Maxwell's
equations as one single octonionic equation
\begin{equation} \label{F=j}
\nabla F = \varrho~.
\end{equation}

As it is clear from (\ref{nabla-F}) and (\ref{current}) in
(\ref{F=j}), as in standard electrodynamics, magnetic current is
absent. This is the result of the ignoring of pseudo-vector and
pseudo-scalar terms in (\ref{current}), and of re-appearance of
these kind of terms in (\ref{nabla-F}) via octonionic products.
Non-associativity of octonions is mainly governed by $j_n$ and
$I$. So the non-existence of magnetic monopoles in classical
electrodynamics can be explained as the use of associativity
limit.


\section{Conclusion and Discussion}

In this paper simple octonionic forms of Dirac's operator and
Maxwell's equations in vacuum were derived. In the classical limit
there is no indication of non-associativity for the electromagnetic
field and derived octonionic Maxwell's equation (\ref{F=j}) is
similar to the bi-quaternion formulations \cite{q-Maxwell}. However,
split octonions that incorporate three vector-like elements should
give a more successful generalization of classical electrodynamics
since non-associativity (which distinguishes octonions from other
normed algebras) also exists in the algebra of Euclidean 3-vectors
used in the classical Maxwell-Heaviside equations. The only new
feature of the octonionic formalism in the approximation used in
this paper is the observation that the non-existence of magnetic
monopoles in classical electrodynamics is connected with the
ignoring of non-associativity. In the case of strong fields the
pseudo-vector and pseudo-scalar parts of octonions can not be
neglected, equations will become more complicated, and we expect to
find new effects in future papers.


\ack
The author would like to acknowledge the hospitality extended
during his visits at the Abdus Salam International Centre for
Theoretical Physics where this work was done.


\end{document}